\documentclass[journal=jacsat,manuscript=article]{achemso}
\usepackage[version=3]{mhchem}
\usepackage{graphicx}
\usepackage{dcolumn}
\usepackage{bm}
\usepackage{float}
\usepackage{booktabs}
\usepackage{amsmath} 

\usepackage{color}
\usepackage{hyperref}
\definecolor{darkred}{rgb}{0.25,0,0}
\definecolor{darkgreen}{rgb}{0,0.25,0}
\definecolor{darkblue}{rgb}{0,0,0.5}
\hypersetup{colorlinks,linkcolor=darkblue,filecolor=black,urlcolor=darkblue,citecolor=darkblue}
\usepackage{multirow}
\usepackage{subcaption}
\usepackage{amsmath}
\usepackage{amssymb}
\usepackage[compact]{titlesec}
\usepackage[utf8]{inputenc}
\DeclareUnicodeCharacter{2212}{-}
\DeclareUnicodeCharacter{0305}{-}

\author{Jianshi Sun}
\affiliation{Institute of Micro/Nano Electromechanical System and Integrated Circuit, College of Mechanical Engineering, Donghua University, Shanghai 201620, China
}%

\author{Shouhang Li}
\email{shouhang.li@dhu.edu.cn}
\affiliation{Institute of Micro/Nano Electromechanical System and Integrated Circuit, College of Mechanical Engineering, Donghua University, Shanghai 201620, China
}%

\author{Zhen Tong}
\affiliation{
School of Advanced Energy, Sun Yat-Sen University, Shenzhen 518107, China
}

\author{Cheng Shao}
\affiliation{
 Thermal Science Research Center, Shandong Institute of Advanced Technology, Jinan, Shandong 250103, China
}

\author{Meng An}
\affiliation{
 Department of Mechanical Engineering, The University of Tokyo, 7-3-1 Hongo, Bunkyo, Tokyo, 113-8656, Japan
}

\author{Xiongfei Zhu}
\affiliation{Institute of Micro/Nano Electromechanical System and Integrated Circuit, College of Mechanical Engineering, Donghua University, Shanghai 201620, China
}%

\author{Chuang Zhang}
\affiliation{Institute of Micro/Nano Electromechanical System and Integrated Circuit, College of Mechanical Engineering, Donghua University, Shanghai 201620, China
}%

\author{Xiangchuan Chen}
\affiliation{Institute of Micro/Nano Electromechanical System and Integrated Circuit, College of Mechanical Engineering, Donghua University, Shanghai 201620, China
}%

\author{Yucheng Xiong}
\affiliation{Institute of Micro/Nano Electromechanical System and Integrated Circuit, College of Mechanical Engineering, Donghua University, Shanghai 201620, China
}%

\author{Thomas Frauenheim}
\affiliation{School of Science, Constructor University, Bremen 28759, Germany
}
\alsoaffiliation{Institute for Advanced Study, Chengdu University, Chengdu 610106, China}

\author{Xiangjun Liu}
\email{xjliu@dhu.edu.cn}
\affiliation{%
 Institute of Micro/Nano Electromechanical System and Integrated Circuit, College of Mechanical Engineering, Donghua University, Shanghai 201620, China
}%

\title{Giant enhancement of hole mobility for 4H-silicon carbide through suppressing interband electron-phonon scattering}
\begin{document}

\clearpage
\begin{abstract}
4H-Silicon Carbide (4H-SiC) possesses a high Baliga figure of merit, making it a promising material for power electronics. However, its applications are limited by its low hole mobility. Herein, we found that the hole mobility of 4H-SiC is mainly limited by the strong interband electron-phonon scattering using mode-level first-principles calculations. Our research indicates that applying compressive strain can reverse the sign of crystal-field splitting and change the ordering of electron bands close to the valence band maximum. Therefore, the interband electron-phonon scattering is severely suppressed, and the out-of-plane hole mobility of 4H-SiC can be enhanced by $\sim200$\% with 2\% uniaxial compressive strain applied. This work provides new insights into the electron transport mechanisms in semiconductors and suggests a strategy to improve hole mobility that could be applied to other semiconductors with hexagonal crystalline geometries.
\end{abstract}

\clearpage

\maketitle
\section{INTRODUCTION}
Silicon Carbide (SiC) has garnered considerable attention in power electronics\cite{elasser2002silicon,eddy2009silicon}, optoelectronics\cite{widmann2019electrical}, and electromagnetics\cite{chen2015mechanical} devices, owing to its unique combination of a high electric breakdown field, high saturation velocity, and excellent thermal conductivity. Among the more than 250 allotropes of SiC, 4H-SiC stands out for its strong physical and chemical stability. However, the extensive utilization of 4H-SiC in applications such as complementary metal-oxide-semiconductor (CMOS)\cite{okamoto2006effect,das2021high} and high-power conversion devices\cite{hull200920} is constrained by its low hole mobility. For instance, the hole mobility of 4H-SiC epilayers on (0001) and slightly off-axis (0001) substrates typically are below 100 $\mathrm{cm\textsuperscript{2}/(Vs)}$ at room temperature\cite{wagner2002aluminum,pernot2005electrical,contreras2017electrical}, which is the bottleneck to the performance of power electronics manufactured by 4H-SiC. Therefore, it is strongly desired to find practical approaches for enhancing the hole mobility in 4H-SiC to optimize the performance of devices and broaden the range of their applications.

Strain engineering has been extensively utilized in advanced logic devices, facilitating the extension of conventional silicon CMOS technologies to at least the 32-nm node.\cite{thompson2006uniaxial,baykan2010strain} In particular, it was reported that applying uniaxial compressive strain along the channel in 45-nm gate-length PMOS devices increases the low field hole mobility by 50\% and the drive current by 25\%.\cite{thompson2004logic} Recently, it was found that strain engineering is an effective way to enhance the hole mobility of gallium nitride (GaN).\cite{ponce2019hole,ponce2019route} Notably, 4H-SiC holds a higher Baliga figure of merit than GaN.\cite{baliga2006silicon} In addition, it is possible to manufacture larger 4H-SiC wafers with fewer defects.\cite{manning2020progress,quast2015high} Therefore, 4H-SiC is more competitive than GaN in power electronics. Furthermore, the existence of fully reversible elastic strain up to $\sim6.2$\% in single-crystal [0001]-oriented 4H-SiC nanopillars has been demonstrated,\cite{fan2021compressive} paving the way for strain modulation of their electrical transport properties. Therefore, it arouses enormous interest in whether the established approach of strain modulation in semiconductors such as Si and GaN can be extended to 4H-SiC.

The electron transport in pristine 4H-SiC is mainly limited by phonon scattering at room temperature and above. Thanks to the development of the theory of first-principles calculation\cite{giustino2017electron}, it is possible to predict the mobility of semiconductors agreeing well with experimental data.\cite{ma2020strain,ma2018intrinsic,ma2020electron,zhou2024isotope,liu2017first,yao2023intrinsic,li2019dimensional,zhu2022giant,ponce2020first,rudra2023reversal,xia2021limits,wang2024electron} It was found that the electron-polar longitudinal-optical (LO) phonon interactions are the main scattering mechanism of 3C-SiC using first-principles calculations.\cite{meng2019phonon} Deng \textit{et al.} accurately predicted drift and Hall mobility of 4H-SiC and they found that the Hall factor is strongly dependent on the direction of the external magnetic field and temperature.\cite{deng2023phonon} Both theoretical and experimental investigations on the effects of strain engineering on the mobility of SiC are lacking and physical mechanisms are yet to be elucidated.

In this work, the microscopic mechanisms underlying the low hole mobility in 4H-SiC are investigated through rigorous mode-level first-principles calculations, and we found strain engineering can overcome the bottleneck of hole mobility. The low hole mobility primarily stems from the strong scattering between holes in the heavy-hole (\textit{hh}) and light-hole (\textit{lh}) bands assisted by acoustic phonons. Additionally, we emphasize that the quadrupole term plays an important role in predicting the electron-phonon scattering rates in 4H-SiC. The out-of-plane hole mobility can be greatly improved by elevating the split-off hole (\textit{sh}) band above both the \textit{hh} and \textit{lh} bands due to reversing the sign of crystal-field splitting under uniaxial compressive strain. In addition, we found that interband electron-phonon scattering is severely suppressed for 4H-SiC with uniaxial compressive strain, and the intraband electron-phonon scattering becomes the dominant scattering mechanism.

\section{THEORY AND METHODS}
Carrier mobility is calculated using the \textit{ab initio} Boltzmann transport equation in the self-energy relaxation time approximation.\cite{ponce2018towards} In this framework, the mobility ($\mu$) can be expressed as
\begin{equation}
    \mu^{\alpha \beta}=\frac{-e}{n_{c} \Omega} \sum_{n} \int \frac{d^{3} \mathbf{k}}{\Omega_{\mathrm{BZ}}} \frac{\partial f_{n \mathbf{k}}}{\partial \varepsilon_{n \mathbf{k}}} v_{n \mathbf{k}}^{\alpha} v_{n \mathbf{k}}^{\beta} \tau_{n \mathbf{k}},
    \label{SE1}
\end{equation}
where $\alpha$ and $\beta$ are the Cartesian coordinates, $n_{\textit{c}}$ represents the carrier concentration, and $\Omega$ denotes the volume of the primitive cell. $\varepsilon_{n \mathbf{k}}$ is the Kohn-Sham energies with electron wavevector $\mathbf{k}$ and band index $n$, $f_{n \mathbf{k}}$ is the Fermi-Dirac distribution and $v_{n \mathbf{k}}$ is the electron band velocity. The carrier-phonon scattering rate is computed within the Fan-Migdal self-energy relaxation approximation\cite{giustino2017electron}
\begin{equation}
    \frac{1}{\tau_{n \mathbf{k}}}=\frac{2 \pi}{\hbar} \sum_{m \mathbf{k}+\mathbf{q}}\left|g_{m n v}(\mathbf{k}, \mathbf{q})\right|^{2} \times\left\{\begin{array}{l}
{\left[n_{\mathbf{q} v}+f_{m \mathbf{k}+\mathbf{q}}\right] \delta\left(\varepsilon_{n \mathbf{k}}-\varepsilon_{m \mathbf{k}+\mathbf{q}}+\hbar \omega_{\mathbf{q} v}\right)} \\
+\left[n_{\mathbf{q} v}+1-f_{m \mathbf{k}+\mathbf{q}}\right] \delta\left(\varepsilon_{n \mathbf{k}}-\varepsilon_{m \mathbf{k}+\mathbf{q}}-\hbar \omega_{\mathbf{q} v}\right)
\end{array}\right\},
    \label{SE2}
\end{equation}
where $\omega_{\mathbf{q}v}$ is the phonon frequency and $n_{\mathbf{q}v}$ is the Bose-Einstein distribution. The electron−phonon matrix element $g_{m n v}(\mathbf{k}, \mathbf{q})=\left(\hbar / 2 \omega_{\mathbf{q} v}\right)^{1 / 2}\left\langle m \mathbf{k}+\mathbf{q}\left|\Delta_{\mathbf{q} v} V\right| n \mathbf{k}\right\rangle$ quantifies the probability amplitude for scattering between the electronic states $n \mathbf{k}$ and $m \mathbf{k}+\mathbf{q}$, with $\Delta_{\mathbf{q} v} V$ the first-order differential of the Kohn−Sham potential with respect to atomic displacement. The Dirac delta functions reflect the conservation of energy during the scattering process. The electron-phonon matrix element contains the long-range part arising from dipolar\cite{verdi2015frohlich}
\begin{equation}
    \begin{aligned}
g_{m n v}^{\mathrm{dip}}(\mathbf{k}, \mathbf{q})= & i \frac{e^{2}}{\Omega \varepsilon_{0}} \sum_{\kappa}\left(\frac{\hbar}{2 \omega_{\mathbf{q} v} M_{\kappa}}\right)^{1 / 2} \sum_{\mathbf{G} \neq-\mathbf{q}} \\
& \times \frac{\left(\mathbf{Z}_{\kappa} \mathbf{e}_{\mathbf{q v}}^{(\kappa)}\right) \cdot(\mathbf{q}+\mathbf{G})}{(\mathbf{q}+\mathbf{G}) \cdot \epsilon \cdot(\mathbf{q}+\mathbf{G})}\left\langle m \mathbf{k}+\mathbf{q}\left|e^{i(\mathbf{q}+\mathbf{G}) \cdot(\mathbf{r}-\tau)}\right| n \mathbf{k}\right\rangle
\end{aligned}
    \label{SE3}
\end{equation}
and quadrupolar interactions\cite{jhalani2020piezoelectric,brunin2020electron}
\begin{equation}
\begin{aligned}
g_{m n v}^{\text {quad}}(\mathbf{k}, \mathbf{q})= & \frac{e^{2}}{\Omega \varepsilon_{0}} \sum_{\kappa}\left(\frac{\hbar}{2 \omega_{\mathbf{q} v} M_{\kappa}}\right)^{1 / 2} \sum_{\mathbf{G} \neq-\mathbf{q}} \frac{1}{2} \\
& \times \frac{(\mathbf{q}+\mathbf{G})_{\alpha}\left(\mathbf{Q}_{\kappa} \mathrm{e}_{\mathbf{q} v, \gamma}^{(\kappa)}\right)(\mathbf{q}+\mathbf{G})_{\beta}}{(\mathbf{q}+\mathbf{G})_{\alpha} \epsilon_{\alpha \beta}(\mathbf{q}+\mathbf{G})_{\beta}}\left\langle m \mathbf{k}+\mathbf{q}\left|e^{i(\mathbf{q}+\mathbf{G})\left(\mathbf{r}-\tau_{\kappa}\right)}\right| n \mathbf{k}\right\rangle,
\end{aligned}
    \label{SE4}
\end{equation}
where $\textit{e}$ is the electron charge and $\epsilon$ is the dielectric tensor. $M_{\kappa}$ and $\tau_{\kappa}$ are the mass and position of the atom with index $\kappa$, respectively. $\mathbf{G}$ are the reciprocal lattice vectors, $\mathbf{Z}_{\kappa}$ is the Born effective charge, $\mathbf{Q}_{\kappa}$ is the dynamical quadrupole tensor, and $\mathbf{e}_{\mathbf{q} v}^{(\kappa)}$ is the phonon eigenvector projected on atom $\kappa$. It should be noted that the $\mathbf{Z}_{\kappa}$ contribution to the quadrupolar interactions is neglected.\cite{ponce2021first}

The Quantum Espresso package\cite{giannozzi2009quantum} is employed for the first-principles calculations with the Perdew−Burke−Ernzerhof (PBE) form of the exchange−correlation functional\cite{perdew2008restoring} and optimized full relativistic norm-conserving pseudopotentials\cite{hamann2013optimized} from PseudoDojo\cite{van2018pseudodojo}. The kinetic energy cutoff for plane waves is set to be 100 Ry, and the convergence of electron energy is set to be $10^{-10}$ Ry. The harmonic force constants are calculated from density-functional perturbation theory (DFPT)\cite{fugallo2013ab,baroni2001phonons}. The electron−phonon scattering rates are calculated using our in-house modified EPW package\cite{ponce2016epw}. The $\mathbf{Q}_{\kappa}$ calculated from linear response\cite{baroni2001phonons,royo2019first} as implemented in ABINIT\cite{gonze2020abinit,romero2020abinit}. The electron−phonon coupling matrix elements are first calculated under coarse $\mathbf{k/q}$-point meshes and then interpolated to dense $\mathbf{k/q}$-point meshes based on maximally localized Wannier functions\cite{marzari2012maximally}. The convergence of out-of-plane hole mobility regarding $\mathbf{k}$-point and $\mathbf{q}$-point meshes at room temperature is verified in [Fig. S1, Supplemental Material]. The accuracy of the PBE exchange-correlation functional is validated by comparing the resulting electronic band structure with the band structure obtained using the screened Heyd-Scuseria-Ernzerhof (HSE) hybrid functional\cite{heyd2003hybrid} (see [Fig. S2, Supplemental Material]). We further computed the out-of-plane hole mobility for carrier concentrations ranging from $10^{11}$ to $10^{15}$ $\mathrm{cm}^{-3}$. The different concentrations are achieved by shifting the Fermi energy with the rigid band approximation.\cite{li2019resolving,sun2024weak} As shown in [Fig. S3, Supplemental Material], the out-of-plane hole mobility is independent of the carrier concentrations within this range. Hence, the carrier concentration for computing the hole mobility is set to $10^{13}$ $\mathrm{cm}^{-3}$ in all cases. More details about the first-principles calculations are provided in the Supplemental Material.

The spin-orbit coupling (SOC) has a pronounced impact on valence bands in the vicinity of the valence band maximum (VBM), while the conduction bands are nearly unaffected, as shown in [Fig. S4, Supplemental Material]. Additionally, the effective masses with and without SOC are calculated using polynomial fitting of the valence band structure near band edges, as shown in [Table S1, Supplemental Material]. It is found that the effective masses of holes are significantly influenced by SOC due to the valence electrons being close to the nucleus\cite{vasileska2017computational}, particularly in the \textit{hh} and \textit{lh} bands along the transverse direction. Therefore, the SOC is included in our valence band structure and hole mobility calculations.

The out-of-plane hole mobilities of 4H-SiC are calculated using energy relaxation time approximation (ERTA)\cite{zhou2016ab}, momentum relaxation time approximation (MRTA)\cite{li2020thermal}, and iterative solution of the Boltzmann transport equation (IBTE)\cite{meng2019phonon,deng2023phonon,ma2018first,ma2022first} are shown in [Fig. S5, Supplemental Material]. The out-of-plane hole mobility predicted by MRTA is very close to that from IBTE in the whole temperature range. However, ERTA tends to underestimate the out-of-plane hole mobility by $\sim28$\% at room temperature, which is consistent with the observations in other polar semiconductors\cite{meng2019phonon,deng2023phonon,ma2018first,ma2022first}. Therefore, MRTA is employed for calculating the hole mobility in the subsequent calculations unless specified.

\section{RESULTS AND DISCUSSIONS}
\begin{figure}[H]
    \centering
    \includegraphics[width=0.95\columnwidth]{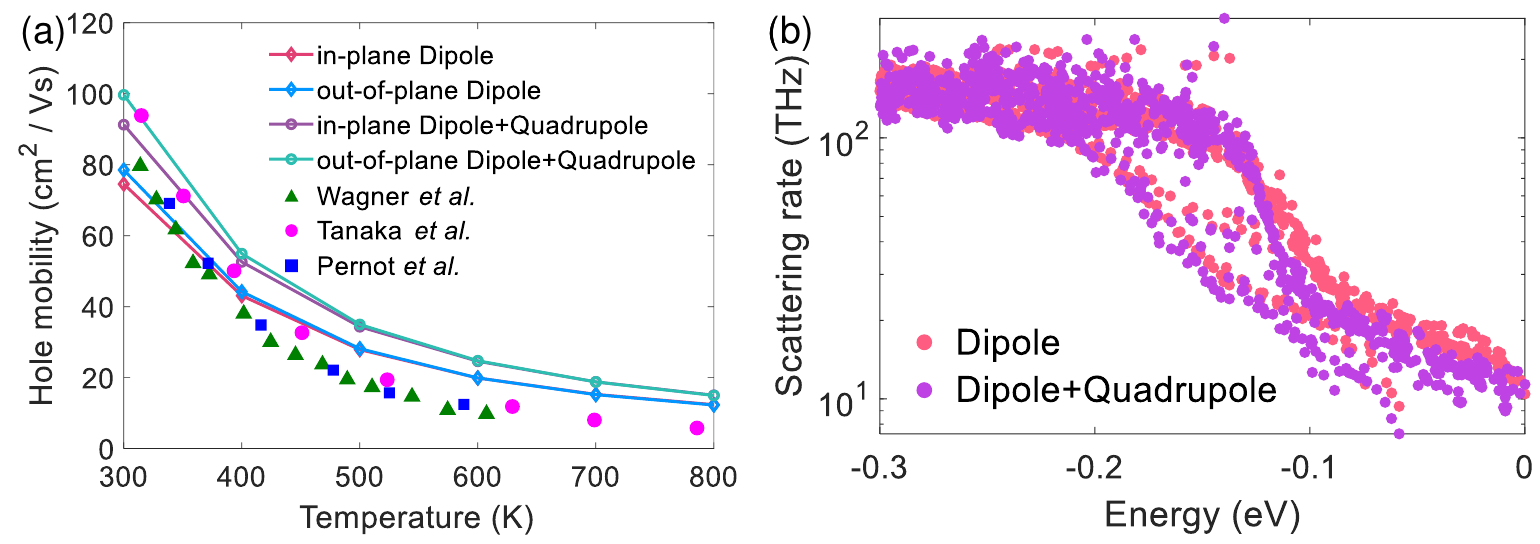}
    \captionof{figure}{(a) The in-plane and out-of-plane hole mobilities with and without quadrupole term as a function of temperature for 4H-SiC. The short-range electron-phonon interactions are included in both cases. The scatters are experimental results reported by Wagner \textit{et al.}\cite{wagner2002aluminum} (triangle), Tanaka \textit{et al.}\cite{tanaka2018theoretical} (circle), and Pernot \textit{et al.}\cite{pernot2005electrical} (square), respectively. (b) The electron-phonon scattering rate with respect to electron energy with and without quadrupole term at room temperature. The electron energy is normalized to the VBM.}
    \label{fig: Figure 1}
\end{figure}

Figure \ref{fig: Figure 1}(a) shows the in-plane and out-of-plane temperature-dependent hole mobilities of 4H-SiC. The short-range electron-phonon interactions are included using Wannier interpolation. The long-range electron-phonon interactions, including both the dipole and quadrupole terms, are calculated. The mobility calculated with the quadrupole term is larger than that only considering the dipole term due to the atomic dynamical quadrupole tensors exhibiting non-zero values. The quadrupole term has main effects on the electron-phonon matrix elements corresponding to acoustic phonons (see [Fig. S6, Supplemental Material]). Electrons are dominantly scattered by acoustic phonons at low temperatures. Therefore, the quadrupole term has a larger correction on the hole mobility at lower temperatures.  However, the quadrupole term leads to a slight modification in the hole mobility at elevated temperatures due to the significant influence of LO phonon scattering induced by the dipole term, together with less important acoustic phonon scatterings. The quantitative differences between our work and Refs. \cite{wagner2002aluminum,pernot2005electrical,tanaka2018theoretical} may be attributed to impurities, dislocations, and other types of point defects in experimental samples. Note that the 4H-SiC is synthesized using the top-seeded solution growth technique, leading to the formation of \textit{n}-type 4H-SiC due to the presence of residual nitrogen.\cite{kusunoki2014nitrogen} Therefore, a substantial amount of hole dopant is required to compensate for its high electron concentration and to achieve \textit{p}-type carrier conduction, with ionized impurity scattering playing a significant role. The effects of ionized impurity scattering on the hole mobility of 4H-SiC are out of the perspective of the present work and will be further discussed in our future work. The influence of the quadrupole term on hole mobility primarily arises from its impact on electron-phonon scattering rates in 4H-SiC. The discrepancy is particularly pronounced for the electron modes within [-0.1, 0] eV normalized to the VBM, which has the dominant contribution to hole mobility. As shown in Figure \ref{fig: Figure 1}(b), the electron-phonon scattering rates with the quadrupole term are smaller than those with only the dipole term taken into account. 

\begin{figure}[H]
    \centering
    \includegraphics[width=0.95\columnwidth]{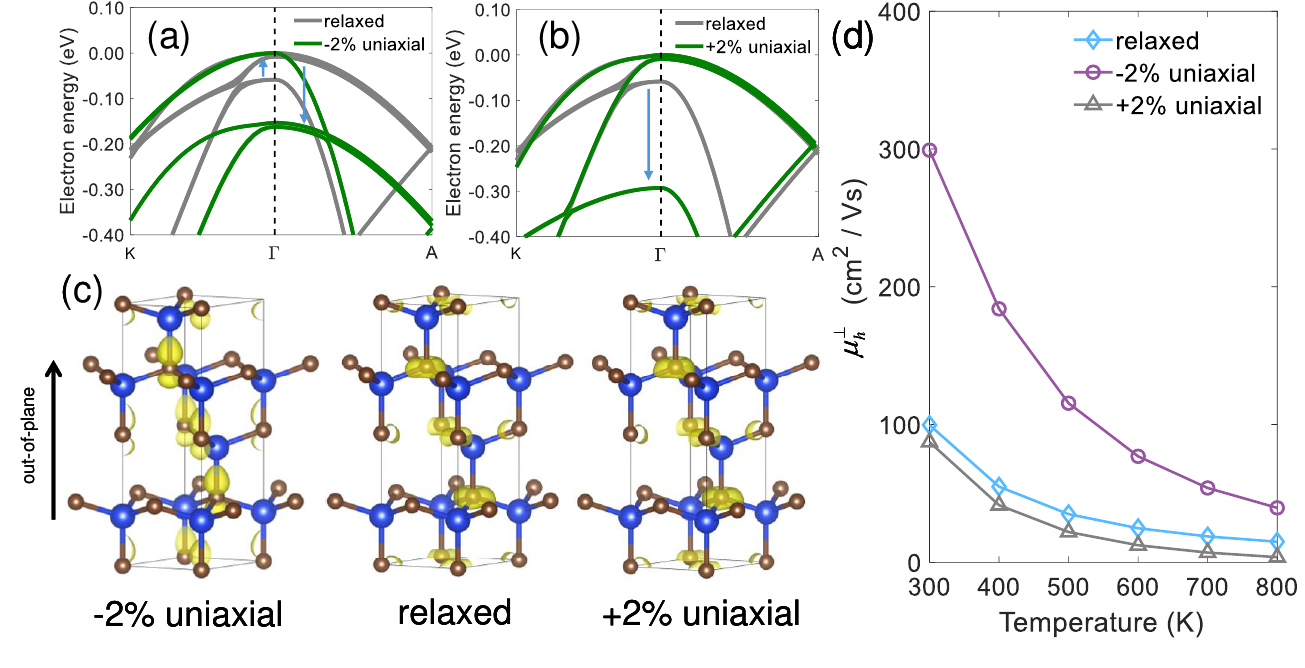}
    \captionof{figure}{The variation in the band structure of 4H-SiC under (a) -2\% and (b) +2\% strain, respectively. The purple arrows represent the shift of the energy bands. The electron energy is normalized to the VBM. (c) Electron wave function characters at the $\Gamma$-point of VBM for relaxed, -2\% strain, and +2\% strain of 4H-SiC, respectively. (d) The out-of-plane hole mobility as a function of temperature for relaxed, -2\% strain, and +2\% strain of 4H-SiC, respectively.}
    \label{fig: Figure 2}
\end{figure}

It should be noted that our predicted values for hole mobilities of the in-plane and out-of-plane 4H-SiC are only 91.3 $\mathrm{cm\textsuperscript{2}/(Vs)}$ and 99.7 $\mathrm{cm\textsuperscript{2}/(Vs)}$ at room temperature, which is significantly smaller than the electron mobility of 4H-SiC, as shown in [Fig. S9, Supplemental Material]. The fundamental reason for the low hole mobility in 4H-SiC lies in the relatively large effective masses of the \textit{hh} and \textit{lh} bands [See Fig. S4, Supplemental Material for bands name]. Naturally, a practical approach to increase hole mobility is to reverse the band ordering of \textit{lh}-\textit{hh} and \textit{sh} bands. As shown in [Fig. S7, Supplemental Material], the \textit{lh} and \textit{hh} bands are separated by the spin-orbit splitting ($\Delta_{\mathrm{so}}$), which is relatively insensitive to uniaxial strain. However, the separation between the \textit{lh}-\textit{hh} and \textit{sh} bands can be manipulated by the crystal-field splitting ($\Delta_{\mathrm{cf}}$)\cite{yan2009strain}, which is very sensitive to uniaxial strain. The $\Delta_{\mathrm{cf}}$ decreases linearly from 241.1 to -158.1 meV for 2\% uniaxial tensile (+2\%) strain to 2\% uniaxial compressive (-2\%) strain. In contrast, the $\Delta_{\mathrm{so}}$ exhibits marginal variation within the same strain range (changing from 13.7 to 13.5 meV). Therefore, it is found that the $\Delta_{\mathrm{cf}}$ reversal will appear with uniaxial strains less than -0.52\%, at which point the \textit{sh} band is shifted above the \textit{lh} and \textit{hh} bands, as shown in Figure \ref{fig: Figure 2}(a) for the case of -2\% strain. When a +2\% strain is applied, the \textit{hh} band has almost no change compared to the unstrained case, while the \textit{lh} and \textit{sh} bands are moved down, particularly the \textit{sh} band exhibiting the most significant change, as shown in Figure \ref{fig: Figure 2}(b). In addition to altering the ordering of the valence band top, the compressive strain can also modify wave function character. For the relaxed 4H-SiC, the wave function character of the VBM is dominated by $\mathrm{C}-p{}_{\mathit{x},\mathit{y}}$ states, and the \textit{sh} band near the zone center mainly consists of $\mathrm{C}-p{}_{\mathit{z}}$ states ([Fig. S8, Supplemental Material]). The wave function character at the VBM changes from the $\mathrm{C}-p{}_{\mathit{x},\mathit{y}}$ states corresponding to the relaxed geometry to the $\mathrm{C}-p{}_{\mathit{z}}$ states for the geometry with -2\% strain. However, the wave function character of the geometry with +2\% strain remains unchanged, as shown in Figure \ref{fig: Figure 2}(c). The effective mass of the \textit{sh} band corresponding to the relaxed geometry along the out-of-plane direction is $m^{\parallel}_{sh}$ = 0.21 $m_e$ at the Brillouin zone center, which is lower than the effective masses of the \textit{hh} ($m^{\parallel}_{hh}$ = 1.62 $m_e$) and \textit{lh} ($m^{\parallel}_{lh}$ = 1.52 $m_e$) bands in the same direction. However, the opposite trend exists for the effective masses in the in-plane direction ($m^{\perp}_{sh}$ = 1.46 $m_e$, $m^{\perp}_{hh}$ = 0.66 $m_e$, and $m^{\perp}_{lh}$ = 0.56 $m_e$). Therefore, the out-of-plane hole mobility of 4H-SiC is expected to significantly increase via reversal of the sign of $\Delta_{\mathrm{cf}}$. Figure \ref{fig: Figure 2}(d) shows the out-of-plane hole mobility of 4H-SiC as a function of temperature for relaxed, -2\% strain, and +2\% strain geometries, respectively. The room-temperature out-of-plane hole mobility of the relaxed geometry is increased by 200\% with -2\% strain applied (from 99.7 to 299.1 $\mathrm{cm\textsuperscript{2}/(Vs)}$). In contrast, the out-of-plane hole mobility is decreased by 12.4\% with +2\% strain applied. Compared to hole mobility, the out-of-plane electron mobility decreases by 41\% under -2\% strain due to a slight increase in the electron effective masses, as shown in [Fig. S9, Supplemental Material]. Given that the out-of-plane electron mobility is still high, we expect uniaxial compressive strain to greatly promote the application of 4H-SiC CMOS devices.

\begin{figure}[H]
    \centering
    \includegraphics[width=0.95\columnwidth]{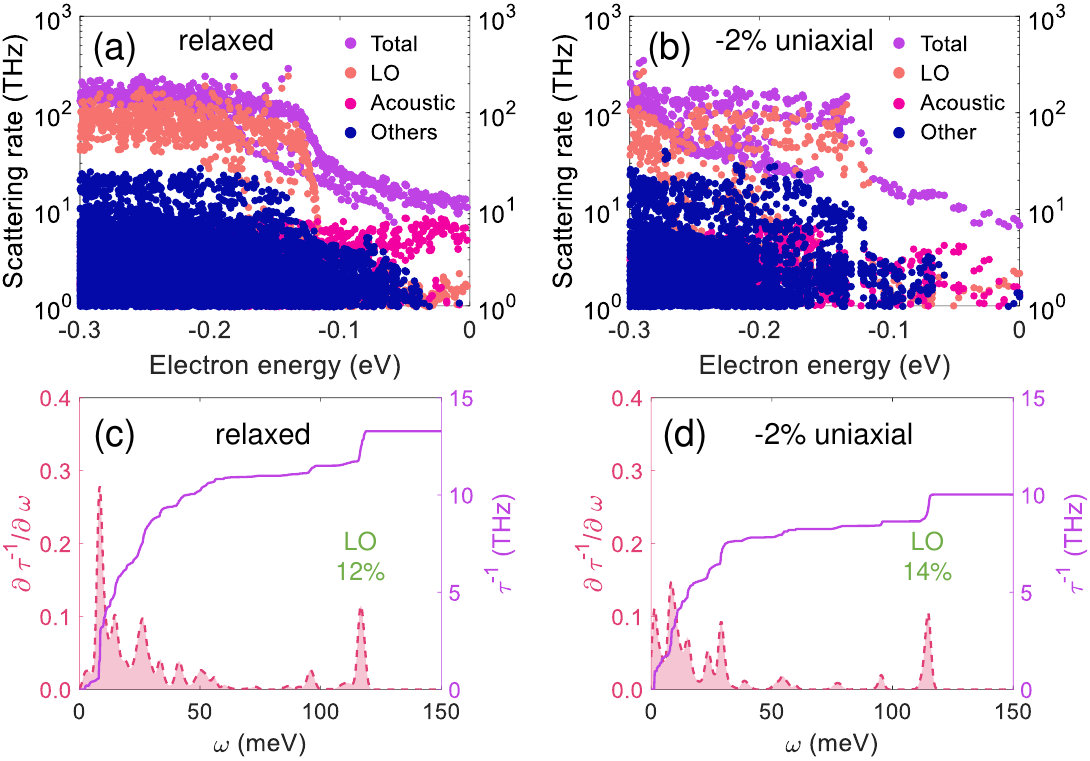}
    \captionof{figure}{Mode-resolved hole scattering rates with respect to the energy below the VBM of (a) relaxed and (b) -2\% strain of 4H-SiC at room temperature. The electron energy is normalized to the VBM. Spectral decomposed hole scattering rates of (c) relaxed and (d) -2\% strain of 4H-SiC as a function of phonon energy calculated 39 meV away from the valence band edges. The peaks represent $\partial \tau^{-1} / \partial \omega$ (left axis), and the blue dashed line represents the cumulative integral $\tau^{-1}$ (right axis). The percentages indicate the cumulative contribution of LO phonons to the total value of $\tau^{-1}$.}
    \label{fig: Figure 3}
\end{figure}

To reveal the underlying mechanisms, the mode-resolved hole scattering rates of relaxed and -2\% strain geometries are shown in Figures \ref{fig: Figure 3}(a) and (b), respectively. For the relaxed case, holes are mainly scattered by acoustic phonons for electron modes in the vicinity of the VBM. For holes with energies lower than 117 meV referred to the VBM, LO phonon scatterings are dominant since the emission of LO phonons becomes available in this region. When a -2\% strain is applied, the total scattering rate decreases significantly. Specifically, the acoustic phonons scattering rates are severely suppressed and their magnitude is comparable to that of LO phonons, as shown in Figure \ref{fig: Figure 3}(b). This can be further validated with the spectral decomposed angular-averaged scattering rate by phonon energy at the most significant carrier energy of 3\textit{$k_b$}\textit{T}/2 (39 meV) away from the VBM, as shown in Figures \ref{fig: Figure 3}(c) and (d). It is found that the main scattering channel comes from acoustic phonons for both the relaxed and -2\% strain geometries. However, the contributions of LO phonons are 12\% and 14\% for the relaxed and -2\% strain cases, which are much lower than the LO phonons contributions in other polar semiconductors\cite{ma2020electron,rudra2023reversal,meng2019phonon,deng2023phonon,ma2018first,ma2022first}.

\begin{figure}[H]
    \centering
    \includegraphics[width=0.95\columnwidth]{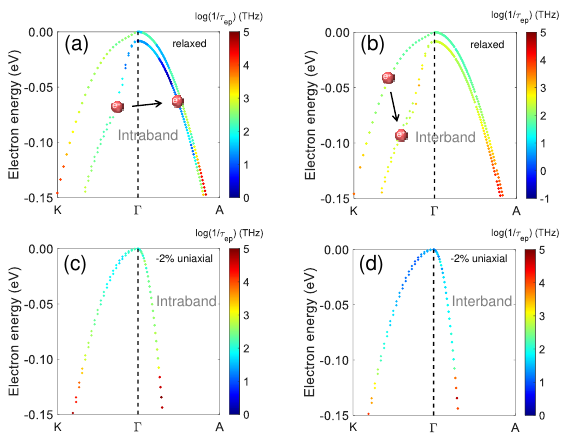}
    \captionof{figure}{Projection of resolved intraband and interband electron-phonon scattering rates at room temperature onto the band structure along the K-$\Gamma$-A high-symmetry path for (a-b) relaxed and (c-d) -2\% strain of 4H-SiC. The electron energy is normalized to the VBM.}
    \label{fig: Figure 4}
\end{figure}

To further reveal the microscopic physical mechanisms for the electron-phonon interactions in 4H-SiC, we project resolved intraband and interband electron-phonon scattering rates onto the band structure for relaxed and -2\% strain geometries, as shown in Figure \ref{fig: Figure 4}. Here, we only present the \textit{hh} and \textit{lh} bands for the relaxed case and the \textit{sh} band for the -2\% strain case since these bands contribute over 95\% to the out-of-plane hole mobility at room temperature ([Tabel S2, Supplemental Material]). Figures \ref{fig: Figure 4}(a) and (b) show that the interband electron-phonon scattering rates are relatively larger than the intraband electron-phonon scattering rates in the relaxed case, particularly for the \textit{lh} band. For the -2\% strain case [Figure \ref{fig: Figure 4}(d)], the increase in curvature of the \textit{sh} band relative to the \textit{hh} and \textit{lh} bands leads to a reduction in the phonon wavevector $\mathbf{q}$ throughout scattering processes, thereby enhancing intraband electron-phonon scattering rates. In contrast, the significant reduction in interband electron-phonon scattering rates is attributed to the \textit{hh} and \textit{lh} bands moving away from the \textit{sh} band, which prohibits energy conservation and decreases the electron-phonon scattering channels. Therefore, the enhancement in the out-of-plane hole mobility of 4H-SiC with uniaxial compressive strains is mainly attributed to the reduction in interband electron-phonon scattering.

Finally, we discuss the practical way to realize the compressive strain in 4H-SiC. Experimentally, 4H-SiC epitaxial layers are typically deposited on thick films, which can lead to strain release through misfit dislocations and cracks in the film. To prevent misfit dislocations and reserve the compressive strain in the sample, it is crucial to grow a properly strained film by keeping its thickness below a critical value \textit{$h_c$}. According to the Fischer model\cite{fischer1994new}, 4H-SiC films can sustain 0.52\% and 2\% compressive strain at thicknesses of 39 nm and 9.4 nm, as shown in ([Fig. S10, Supplemental Material]). Therefore, the thickness of 4H-SiC films needs to be meticulously controlled to prevent the enhancement of hole mobility from being suppressed by dislocation scattering.

\section{CONCLUSIONS}
In summary, the hole mobility of 4H-SiC geometries with and without strains is investigated through mode-level first-principles calculations. It is found that the origin of the low hole mobility lies in the large effective masses of heavy and light hole bands, as well as strong interband scattering mediated by low-energy acoustic phonons. Additionally, the quadrupole corrections for electron-phonon interactions are crucial for accurately predicting electronic transport properties in 4H-SiC. We proposed that the out-of-plane hole mobility can be greatly enhanced by modifying the ordering of electron bands in the vicinity of the valence band maximum and significantly suppressing the interband electron-phonon scattering. The band inversion phenomenon can be achieved through strain engineering to reverse the sign of crystal-field splitting. This work provides profound insights into electron transport in 4H-SiC that deserve experimental verification to expedite advancements in SiC-based CMOS technology.

\begin{acknowledgement}
S.L. was supported by the National Natural Science Foundation of China (Grant No. 12304039), the Shanghai Municipal Natural Science Foundation (Grant No. 22YF1400100), the Fundamental Research Funds for the Central Universities (Grant No. 2232022D-22), and the startup funding for youth faculty from the College of Mechanical Engineering of Donghua University. X.L. was supported by the Shanghai Municipal Natural Science Foundation (Grant No. 21TS1401500) and the National Natural Science Foundation of China (Grants No. 52150610495 and No. 12374027). The computational resources utilized in this research were provided by the Shanghai Supercomputer Center and the National Supercomputing Center in Shenzhen.

\end{acknowledgement}

\begin{suppinfo}
Detailed information regarding computational details, verification of first-principles calculations, PBE and HSE electronic band structures, carrier concentration impact on hole mobility, band structure with and without spin-orbit coupling, comparison of hole mobility prediction methods, mode hole scattering rates without and with quadrupole term, crystal-field splitting reversal under uniaxial strain, projection of $\mathrm{C}-p$ orbitals onto the valence band structure, electron mobility with and without compressive strain, critical thickness of 4H-SiC thin film, hole effective masses with and without SOC, band contributions to out-of-plane hole mobility.
\end{suppinfo}

\bibliography{bibliography}

\end{document}